\documentclass[conference]{IEEEtran}
\usepackage{algorithm}
\usepackage{algorithmic}
\usepackage{graphicx}
\usepackage{caption}
\usepackage{multirow}
\usepackage{tabularx}
\usepackage{adjustbox}
\makeatletter
\newcommand\fs@norules{\def\@fs@cfont{\bfseries}\let\@fs@capt\floatc@ruled
  \def\@fs@pre{}%
  \def\@fs@post{}%
  \def\@fs@mid{\kern3pt}%
  \let\@fs@iftopcapt\iftrue}
\makeatother
\floatstyle{norules}
\restylefloat{algorithm}

\begin{document}

\title{Spam filtering on forums: A synthetic oversampling based approach for imbalanced data classification}

\author{\IEEEauthorblockN{Pratik Ratadiya}
\IEEEauthorblockA{Department of Computer Engineering\\
Pune Institute of Computer Technology, \\
Maharashtra, India. \\
Email: prratadiya@gmail.com}
\and
\IEEEauthorblockN{Rahul Moorthy}
\IEEEauthorblockA{Department of Computer Engineering\\
Pune Institute of Computer Technology, \\
Maharashtra, India. \\
Email: rahulmoorthy9.6@gmail.com}}

% make the title area
\maketitle

% creates the second title. It will be ignored for other modes.
\IEEEpeerreviewmaketitle

 \begin{abstract}
     Forums play an important role in providing a platform for community interaction and grievance redressal. The introduction of irrelevant content or spam by some individuals for commercial and social gains tends to degrade the professional experience presented to the users of the forum. Automation in moderating the relevancy of the posted content is highly desired. Machine learning has been the cornerstone for text classification and finds applications in spam email detection, fraudulent transaction detection, text segmentation etc. The balance of positive and negative classes in training data is essential in the case of classification algorithms to make the learning task more efficient and accurate. However, in the case of forums, the spam content is usually sparse compared to the relevant content giving rise to a bias towards the latter in the training process. A model trained on such data will fail to classify a spam sample due to the present bias. An approach based on Synthetic Minority Over-sampling Technique(SMOTE) is presented in this paper to solve the problem of imbalanced training data. The presented approach involves synthetically creating new minority class samples from the existing ones till the balance in data is acheived. Training data is oversampled and then passed through various classification algorithms for which the performance over different metrics is recorded. The results were analyzed on the data of forums of Spoken Tutorial, IIT Bombay over different standard performance metrics and revealed that models trained after Synthetic Minority oversampling outperform the ones trained on imbalanced data by substantial margins. An empirical comparison of the results obtained by both SMOTE and without SMOTE for various supervised classification algorithms have been presented in this paper. Synthetic oversampling proves to be a critical technique for achieving uniform class distribution which in turn yields commendable results in text classification. The presented approach can be further extended to content categorization on educational websites thus helping to improve the overall digital learning experience.
\end{abstract} 

\section{Introduction}
% no \IEEEPARstart
 The intention of forums is to provide a platform between the users and administration for resolving issues and grievance redressal. However, there are instances of users posting malicious content and links on these platforms which degrades the core objective of the website. Currently, there is manual moderation for quality check to remove any irrelevant content. Such moderation increases the cost as well as the time required for completion of the objective which can be reduced by automating this process. \par
 There has been ample research in recent years to develop automated spam classification systems using supervised machine learning algorithms. Awad and Elseuofi applied various machine learning techniques for automated classification of emails \cite{spam1}. Manlangit \textit{et al.} proposed an intelligent system which could detect fraud transactions using classification algorithms \cite{spam2}. These systems made use of labelled data to plot a decision boundary based on which a new test case was classified into the appropriate class. \par 
 One of the major issues faced for developing such classifier systems for forums is the imbalance in data. Majority of the content posted on forums is in accordance with the intended goal(non spam). As a result, the frequency of spam samples is sparse in comparison to the non spam samples. Conventional Machine Learning algorithms when trained on such data will tend to have a bias towards the majority class which will lead to a new spam sample being accepted as relevant. This inconsistency can be resolved by feeding balanced data to the algorithm. The balance amongst classes can be achieved by either undersampling the majority class or oversampling the minority class. In 2008, Tang \textit{et al.} presented a hybrid intelligent system based on undersampling of the majority class which could analyze the behaviour of sender \cite{undersample}. However, undersampling of data poses the risk of losing crucial information especially in cases where the number of minority class samples is extremely small, thus affecting the overall performance of the system. In this paper we propose an approach which involves synthetically oversampling the minority class to solve the imbalance problem. This increases the training size so as to bring uniformity in the data by reducing variation in distribution of classes. This in turn enhances the results achieved by the system. \par
 The rest of the paper is as follows: Section 2 provides an overview of the pre-processing techniques. The proposed approach is described in section 3. Description of the dataset, evaluation metrics and experimental results are provided in section 4. Finally, section 5 draws the conclusion and suggests future direction of the work.

\section{Preprocessing techniques}
The data obtained from the forums is in text format and cannot be feeded directly to the machine learning algorithm. Certain preprocessing techniques need to be applied on the data which help in extracting useful information. The text data is converted into feature vectors which are then passed to the classification algorithm. The preprocessing techniques include:
\subsection{Stripping of HTML tags}
The text is removed of any html tags and scripts as they are not relevant to the intent of the message. Apart from this special characters like —\&nbsp;,newline and tab characters are also removed. Words with less than 3 characters are removed as they do not provide relevant information and tend to divert the model function.
\subsection{Stop words}
Commonly used words like ’are’,’is’,’they’,’this’ etc can be found in both spam as well as non-spam content and thus cannot be a deciding factor for classification. Such words are called stop words and we remove them from our text.
\subsection{Vectorization}
The filtered text is now converted into a sparse matrix of tokens by vectorizing them. The various vectorizer techniques include count vectorizer and Tf-idf vectorizer. Tf-Idf(Term
frequency inverse document frequency) vectorizer is used by us in this project. The mathematical equation for Tf-Idf is as follows:
\begin{center}
\begin{equation}
    tfidf_{i,d} = \frac{n_{i,j}}{\sum_k n_{k,j}} \cdot \log{\frac{|D|}{|{d : t_i \in d}|}}
\end{equation}

\end{center}
The Tf-Idf weight is composed by two terms: the first computes the normalized Term Frequency (TF), aka. the number of times a word appears in a sample, divided by the total number of words in that sample; the second term is the Inverse Document Frequency (IDF), computed as the logarithm of the number of the samples in the dataset divided by the number of samples where the specific word appears. \par
The above mentioned preprocessing techniques are implemented using the beautiful soup and scikit learn packages in Python 3.6. After carrying out these techniques, the text data is now converted to a sparse matrix consisting of numbers which can be fed to a classification algorithm. As a part of the evaluation process, we split the data into train and test sets with the model being trained on 80\% of the data and results are calculated for the remaining 20\% of dataset. For any new test case, the above preprocessing methods still need to be carried out and the class label is predicted for the feature vector of the test case.
% conference papers do not normally have an appendix

\section{Proposed approach}
The training dataset initially contains uneven distribution of classes. We propose oversampling the minority class using Synthetic Minority Oversampling Technique(SMOTE) to reduce the variance present in the training dataset. SMOTE was first introduced by Chawla \textit{et al.} in 2002 \cite{smotpaper}. The technique involves creation of 'synthetic' samples rather than replacement techniques which are traditionally used for oversampling. The algorithm operates on the feature vectors created earlier by introducing artificial samples along the line joining all of the k minority class nearest neighbors for every minority class sample. It is generally implemented in the training phase so as to remove bias by balancing the data. Synthetically oversampling the test data does not enhance the performance of the classification algorithm. We implement the technique on training data such that the number of minority and majority class samples become equal. 

 \begin{algorithm}[!ht]
 \caption{Algorithm for SMOTE} 
 \begin{algorithmic}[1]
 \renewcommand{\algorithmicrequire}{\textbf{Input:}}
 \renewcommand{\algorithmicensure}{\textbf{Output:}}
 \REQUIRE Number of minority class samples T; Number of majority class samples M; Number of nearest neighbors k
 \ENSURE  M minority class samples
  \FOR {$i = $1 to $T$}
  \STATE Compute k nearest neighbors for i
  \WHILE{T != M}
  \STATE Randomly chose one of the k nearest neighbor of i in feature space, say nn
  \STATE Compute vector between nn and i and multiply it with a random number between 0 and 1
  \STATE Synthetic sample = i + computed vector  
  \ENDWHILE
  \ENDFOR
 \RETURN $T$
 \end{algorithmic}
 \end{algorithm}
 
 After passing the feature vectors through this algorithm, there is modification in the feature space as observed through figure 1 and 2. It should be noted that the samples in blue indicate the majority class which are comparatively more abundant than the minority class(green) in fig 1. The increase in minority class samples after implementing SMOTE is visible in figure 2. We can also infer that the formation of decision boundary to distinguish the two classes in feature space is now comparatively easier, thus enhancing the performance of the classification algorithm. 
 
\begin{figure}[H]
\centering
\includegraphics[width=2.5in]{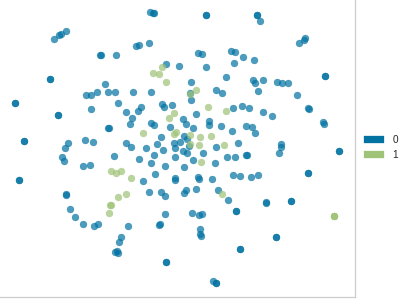}
\caption{Feature space of training vector before applying SMOTE}
\label{fig_sim}
\end{figure}

\begin{figure}[H]
\centering
\includegraphics[width=2.5in]{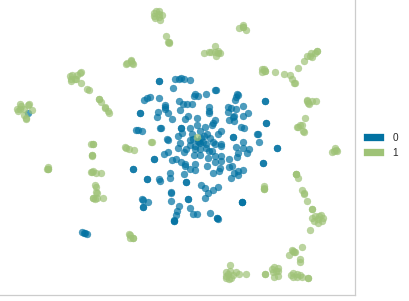}
\caption{Feature space of training vector after applying SMOTE}
\label{fig_sim2}
\end{figure}
The balanced training dataset is now passed through a supervised classification algorithm. The task of the algorithm is to plot a decision boundary which fits the training data such that it precisely partitions the underlying vector space into two sets, one for each class. For a new test case, the previously mentioned preprocessing steps are carried out and its feature vector is plotted in the feature space. The classifier will classify it as the class which is on the same side of the decision boundary as the vector.\par
We implement the SMOTE algorithm in Python using the imblearn package. The various classification algorithms are implemented using the scikit learn library \cite{skl1,skl2}. The pandas and matplotlib packages are used for reading and visualization of the data respectively.
% use section* for acknowledgement

\section{Evaluation}
\subsection{Dataset description}
The dataset for this task was gathered from the forums of Spoken Tutorials project, IIT Bombay. Spoken tutorial project provides tutorials on FOSS available in several Indian languages for the learners\cite {spoken}. A forum has been setup to address problems of students as well as contributors. The data was collected from the forums which consisted of questions asked by users and the replies given. A label was assigned to each case. Label 0 indicated 'non spam' or relevant content whereas the label 1 indicated spam content. The dataset comprised of 313 samples-273 non spam samples and 40 spam samples. It was then split into training and testing dataset with the training set consisting of 201 non spam samples and 33 spam samples. The distribution of training data before and after applying SMOTE is shown in table 1.\newline

\begin{table}[b]
% increase table row spacing, adjust to taste
\def\arraystretch{2}
\centering

\begin{tabularx}{\columnwidth}{|l|X|X|}
\hline
& Number of majority class samples & Number of minority class samples\\
\hline
Initial stage & 201 & 33\\
\hline
After applying SMOTE & 201 & 201\\
\hline
\end{tabularx}

\caption{Distribution of training dataset}
\end{table}

\subsection{Performance metric}
Accuracy is usually considered as the performance metric for classification algorithms. It is the ratio of correctly detected samples to the total number of test samples. However, in case of imbalanced datasets the accuracy will provide a wrong perception of the results. For example, a dataset consisting of 95 non-spam and 5 spam samples will produce an accuracy of 95\% by predicting all samples as non-spam. The model has failed to recognize the spam class samples and is thus deemed faulty. The performance metrics which provide a better interpretation in such cases are: 

\begin{center}
    Precision = $\frac{\displaystyle TP}{\displaystyle TP + FP}$ \newline\newline
    Recall = $\frac{\displaystyle TP}{\displaystyle TP + FN}$
    \newline\newline
    F1 Score = 2 * $\frac{\displaystyle Precision * Recall}{\displaystyle Precision + Recall}$
\end{center}
where\newline TP: True positive(no. of outcome where the model correctly predicts the positive class)\newline
FP: False positive(outcome where the model incorrectly predicts the positive class)\newline
FN: False negative(outcome where the model incorrectly predicts the negative class)
The reason is that these metrics are more focused on the positive class(Spam) than on the negative class and actually measure the probability of correctly detecting positive values which is crucial in case of spam classification.

\subsection{Evaluation of results}
The enhanced training dataset is passed through various classification algorithms namely: Multinomial Naive Bayes, Logistic regression, Linear SVC(Support vector clustering) and decision tree. The performance of each algorithm based on the above mentioned metrics is tabulated in table no 2.

\begin{figure}[b]
\centering
\includegraphics[width=3.5in]{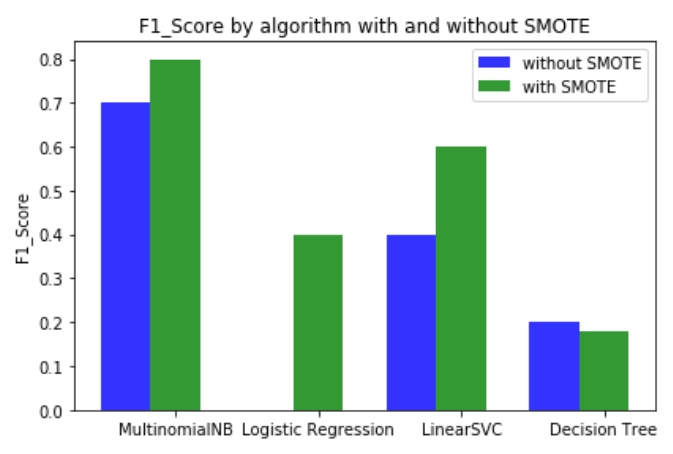}
\caption{F1 Score metric representation}
\label{fig_sim3}
\end{figure}

\begin{table*}[t]

\def\arraystretch{1.5}
\begin{tabular}{|l|c|c|c|c|c|c|c|c|}
\hline
\multirow{2}{*}{Metric} & \multicolumn{2}{c|}{Multinomial NB}                                  & \multicolumn{2}{c|}{Logistic Regression}                             & \multicolumn{2}{c|}{Linear SVC}                                      & \multicolumn{2}{c|}{Decision Tree}                                   \\ \cline{2-9} 
                        & \multicolumn{1}{l|}{With SMOTE} & \multicolumn{1}{l|}{Without SMOTE} & \multicolumn{1}{r|}{With SMOTE} & \multicolumn{1}{l|}{Without SMOTE} & \multicolumn{1}{l|}{With SMOTE} & \multicolumn{1}{l|}{Without SMOTE} & \multicolumn{1}{l|}{With SMOTE} & \multicolumn{1}{l|}{Without SMOTE} \\ \hline
Accuracy                & 0.96                               & 0.95                                  & 0.92                               & 0.91                                  & 0.95                               & 0.936                                  & 0.88                               & 0.89                                  \\ \hline
Precision               & 1.0                               & 0.71                                 & 0.28                              & 0.0                                 & 0.57                              & 0.28                                 & 0.14                              & 0.14                                 \\ \hline
Recall                  & 0.7                              & 0.71                                 & 0.6                              & 0.0                                 & 0.8                              & 1.0                                 & 0.25                              & 0.3                                 \\ \hline
F1 Score                & 0.82                              & 0.71                                 & 0.4                              & 0.0                                 & 0.6                              & 0.4                                 & 0.18                              & 0.2                                 \\ \hline
\end{tabular}

\caption{Performance of various classification algorithms on Spoken Tutorial forum dataset}
\end{table*}

\begin{figure}[H]
\centering
\includegraphics[width=3.5in]{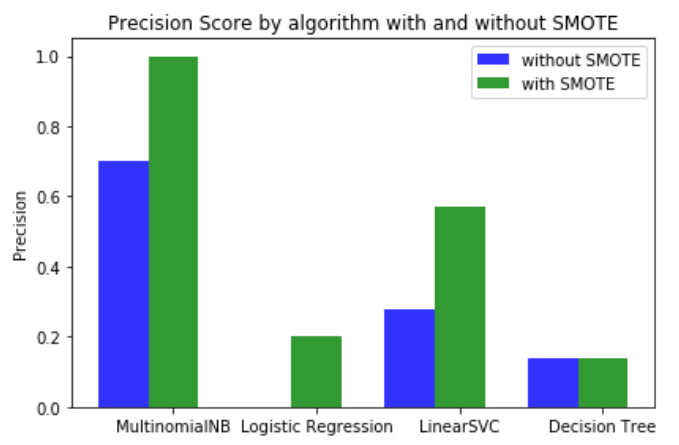}
\caption{Precision Score metric representation}
\label{fig_sim4}
\end{figure}
 It can be observed that applying SMOTE on training data produces better metric performance compared to models trained on imbalanced data for major classification algorithms.
\section{Conclusion}
We have observed that algorithms working with SMOTE outperforms algorithms trained on imbalanced data with margins as high as 10\%. The effect of uneven class representation is negated by this technique. It can thus prove to be extremely useful for classification of cases in forums. Future work includes modifications in the SMOTE function for better generalization of the minority class. The performance of the technique can be enhanced for decision tree based algorithms through parameter tuning. The technique can also be meshed with deep learning algorithms to enhance the classification result. The idea can be further extended to create automated tagger for contents posted on educational websites which would improve content organization.
\section*{Acknowledgment}

 We would like to thank Spoken Tutorial Project, IIT Bombay for their assistance in the collection of data used in this project. We would also like to extend our thanks to Prof. Kannan Moudgalya, IIT Bombay for his encouragement and useful critiques for this research work.

% that's all folks
\end{document}